\documentclass{sig-alternate}
\usepackage[latin1]{inputenc}
\usepackage{amsmath}
\usepackage{amsfonts}
\usepackage{amssymb}
\usepackage{graphicx}
\usepackage{multicol}
\usepackage[usenames]{color}
\usepackage{url}

\begin{document}
%
\conferenceinfo{Wns3 2011\,}{March 25, Barcelona, Spain.}   
\CopyrightYear{2011} 
\crdata{}  

\title{A Multipath TCP model for ns-3 simulator}
%
%
%
%
%

\numberofauthors{2} 
%
\author{
%
%
\alignauthor
Bachir Chihani\\
       \affaddr{Orange Labs}\\
       \affaddr{Sophia Antipolis, France}\\
       \email{bachir.chihani@orange-ftgroup.com}
\alignauthor
Denis Collange\\
       \affaddr{Orange Labs}\\
       \affaddr{Sophia Antipolis, France}\\
       \email{denis.collange@orange-ftgroup.com}
}

\maketitle
\begin{abstract}

We present an implementation of Multipath TCP (MPTCP) under the NS-3 open source network simulator. MPTCP is a promising extension of TCP currently considered by the recent eponymous IETF working group, with the objective of improving the performance of TCP, especially its robustness to variable network conditions. We describe this new protocol, its main functions and our implementation in NS-3. Besides this implementation compliant to the current versions of the IETF drafts, we have also added and compared various packet reordering mechanisms. We indeed notice that such mechanisms highly improve the performance of MPTCP. We believe that our implementation could be useful for future works in MPTCP performance evaluation, especially to compare packet reordering algorithms or coupling congestion control mechanisms between subflows. 

\end{abstract}




\section{Introduction}
Nowadays mobile equipment have often more than one single network interface. For instance, laptops have usually at least both a wired (Ethernet) and a wireless (Wifi) network adapters. Similarly smartphones and tablet PCs can reach the Internet either through Wifi or through a cellular network (UMTS or 3G+). 

Another fact is that network operators usually duplicate links and equipments in order to protect their networks against failures, especially in the access and the backhaul networks. 
Moreover the backbone networks are generally meshed.
In this context many paths may exist between any two endpoints. The idea to use concurrently many paths has then emerged, to improve the robustness and performance of end-to-end connections. Such multipath connections can indeed balance the load between the different paths, switching dynamically and automatically the traffic from congested, disrupted or broken links to the best paths.  

A lot of studies have considered the implementation of multipath capabilities at different layers: at the application layer \cite{hacker02}, at the transport layer \cite{ong02}, \cite{hasegawa05}, \cite{dong07}, \cite{sarkar06}, \cite{zhang04}, \cite{rojviboonchai02}, \cite{hasegawa05}, etc.
Following these last references, and \cite{wischik08}, we think that the transport layer is a good place to implement multipath capabilities.

At this layer, end-systems can gather information about each used path: capacity, latency, congestion state. These information can then be used to react to congestion events in the network by moving the traffic away from congested paths.
An IETF's working group has recently been created to specify a multipath protocol at the transport layer Multipath TCP \cite{ford10} (MPTCP). More precisely, this working group develop protocol extensions to Transmission Control Protocol (TCP), the most used transport protocol on Internet, to handle multiple sub-connections following different paths between two endpoints. 

Previous works in simulating MPTCP \cite{htsim} focused on congestion control mechanisms without implementing other parts of MPTCP. This restriction was inherited from the used simulator, which does not allow the creation of an accurate model of the simulated protocol. Our contribution consists of the proposal of a model of MPCTP with a better fidelity, and also the proposal of enhancing the current version of MPTCP with packet reordering mechanisms to cope with the variety of path characteristics.    

In this paper we first describe MPTCP in section \ref{sec:drafts}, as it is specified in the current versions of the IETF drafts \cite{ford10}. We describe in section \ref{sec:reordering} the implementation of two packet reordering algorithms, and how to adapt them to MPTCP.
Then we present in section \ref{sec:implementation} our implementation of MPTCP under NS-3, conform to these IETF drafts, and with some of the congestion control algorithms proposed for MPTCP, for example in \cite{raiciu09}. 
In section \ref{sec:simulation} we describe the simulation scenario, and comment the behaviour of the congestion window according to the used packet reordering algorithm.
Finally, in section \ref{sec:conclusion} we conclude the paper.

\section{Multipath TCP}\label{sec:drafts}

An IETF working group has recently been created to specify a multipath protocol for the transport layer. They propose MPTCP \cite{ford10} (Multipath TCP), an extension of TCP to handle multiple paths  between two endpoints. MPTCP is designed with three major goals:
\begin{enumerate}
\item 	{\bf Improve throughput}: the performance of a multi-path flow should be at least as good as this of a single-path flow on the best route.
\item 	{\bf Do no harm}: a multi-path flow should not take up any more capacity on any one of its paths than a single-path flow using that route.
\item 	{\bf Balance congestion}: a multi-path flow should move as much traffic as possible away from the most congested paths.
\end{enumerate}

\subsection{Main mechanisms}

With MPTCP, the transport layer is splitted into two sublayers. The upper one gathers the functionalities for connection management (establishing connection, reordering packets, etc.). The lower sublayer manages a set of subflows that can be seen each as one single TCP flow.
MPTCP distinguishes two spaces of sequence numbers, one for each sublayer. Each subfow has its own sequence space which is similar to the Standard TCP sequence number, identifying bytes within a subflow. At the connection level, another sequence space is used to reorder the TCP segments before sending them to the application layer.

The MPTCP protocol uses new TCP options to exchange signalling information between peers: especially: 
\begin{description}
\item[{\bf MPC}] (Multipath Capable) is used during the three-way handshake to establish a multipath TCP connection.  
\item[{\bf DATA}] {\bf FIN} is used to inform the remote peer of the end of data and to close the multipath TCP connection.
\item[{\bf ADD}] and {\bf REMOVE} Address (IPv4) are used to inform the remote peer of the availability of a new address or to ask it to ignore an existing one.
\item[{\bf JOIN}] is used to initiate a new sub-flow (packet flow on a route) between a not already used couple of addresses.
\item[{\bf DSN}] (Data Sequence Number) is used as a map between the subflow level and the data sequence space number.
\end{description}

\subsubsection{Connection establishment}

\begin{figure}[h]
\centering
\includegraphics[scale=0.5]{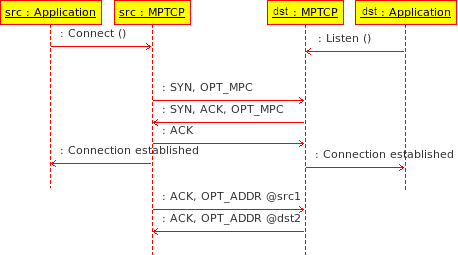} 
\caption{MPTCP connection establishment}
\label{ConEstab}
\end{figure}

Figure \ref{ConEstab} illustrates the process of establishment of a MPTCP connection. After that the source application sends a Connect() call, the transport layer establishes a connection with the destination peer which was waiting for receiving connection requests. The establishment is TCP-like (three way handshake) with the use of MPC option to inform the callee that the initiator is able to exchange data using Multipath TCP. To initiate a new subflow, the peers must first exchange their additional IP addresses. The current MPTCP draft do not specify how the exchange may happen. We have chosen to send additional TCP segments. These segments handle the ADDR (Add address) option and they are sent just after the successful establishment of the MPTCP connection.

\subsubsection{Subflow initiation}

\begin{figure}[h]
\centering
\includegraphics[scale=0.5]{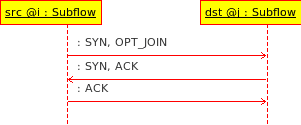} 
\caption{MPTCP sub-flow initiation}
\label{SubflowInit}
\end{figure}

Figure \ref{SubflowInit} shows the initiation of a new subflow and the presence of a JOIN in a SYN segment. To maximize the chance that the subflow under initiation takes a path which is disjoined with previously established paths, each IP address is only used by one subflow.

\subsection{Traffic control}

MPTCP redefines some TCP mechanisms so that they fit the multipath context. 
Congestion control allows the sender to regulate its throughput according to the available network resources. 
With MPTCP, the congestion control is performed at the subflow level. Each subflow has its own congestion window. However the congestion windows of the different subflows of a given TCP connection may be coupled to improve its performance. 
Besides, at the upper sublayer, the MPTCP receiver has a  single global receiving window shared between the set of the established subflows. The objective is to do not limit the speed of some subflows.
Four different algorithms have been proposed by Raiciu et al. in \cite{raiciu09}, coupling in various ways the congestion windows of active subflows: Uncoupled, Fully Coupled, Linked Increase, and RTT Compensator. They consider a simple extension of the standard congestion control mechanism TCP Reno  in case the round-trip time is the same for all the available paths $r = 1,..., N$. 

With the algorithm Uncoupled, the congestion window of each subflow behaves like for a single Standard TCP connection. 
Let $w_r$ be the congestion window on path $r$, and $w = \sum_r w_r$

Algorithm {\bf Fully Coupled}
\begin{itemize}
\item $w_r = w_r + \frac{1}{w}$ per ACK on path $r$
\item $w_r = max(w_r - \frac{w}{2}, 1)$ per loss event on path $r$
\end{itemize}

Most of the time either one path or another is used with this algorithm, and rarely both. This phenomenon is called "flappiness". To reduce this flappiness, the authors proposed the following algorithm:

Algorithm {\bf Linked Increases} \cite{raiciu10}
\begin{itemize}
\item $w_r = w_r + \frac{a}{w}$ per ACK on path $r$
\item $w_r = \frac{w_r}{2}$ per loss event on path $r$
\end{itemize}

In more general case when the round-trip times are not equal for the all paths, the authors adjust the precedent algorithm:

Algorithm {\bf RTT Compensator}
\begin{itemize}
\item $w_r = w_r + min(\frac{a}{w}, \frac{1}{w})$ per ACK on path $r$
\item $w_r = \frac{w_r}{2}$ per loss event on path $r$
\end{itemize}

\section{Packet reordering}\label{sec:reordering}
With Standard TCP, in networks with large packet jitter, i.e. when the end-to-end delay may vary a lot, packets may arrive out-of-sequence. This may for example be the case on wireless networks where mobile devices may change the used hotspot for accessing the Internet. The reordering of a packet makes the receiver responding with duplicated acknowledgements, and this may induce the sender to infer wrongly a packet loss. To avoid this problem and to distinguish clearly between packet losses due to congestion in the network and reordering due to transmission jitter, many mechanisms have been proposed for TCP. Leung et al. cite some of such mechanisms in \cite{leung07}. Some of these mechanisms have been specified by the IETF \cite{EifelRFC} \cite{DsackRFC} \cite{FrtoRFC}, and some of them are implemented in the operating systems \cite{FrtoRFC} (especially in the most frequent like Windows, Linux...) where they may be active by default, or not. 

In multipath context, packets may also arrive out-of-sequence as the different paths may have different characteristics (especially the end-to-end delay), or congestion state (and then different queuing delays). The out-of-sequence arrival will create a problem for MPTCP while re-assembling packets at the connection level, and not at the subflow level because subflows are independent. So we suggest to consider for MPTCP similar reordering mechanisms than those proposed for Standard TCP on wireless networks. 

Some of the algorithms, like Eifel \cite{EifelRFC} \cite{Eifel} and DSACK \cite{DsackRFC}, need to store the connection state (the congestion window cwnd, the slow-start threshold ssthresh, etc.) before retransmitting. When a spurious retransmission is detected, the saved state is restored. Other algorithms, like the Blanton-Allman algorithm \cite{leung07} and the RR-TCP (Reordering-robust TCP) \cite{leung07}, adjust the threshold dupthresh which is the number of received duplicated acknowledgement after which the sender considers a packet as lost and retransmits it. Others proposed mechanisms make the receiver delaying the transmission of duplicated ACK, like Paxson algorithm \cite{leung07}, or make the sender delaying its response to congestion (reception of three duplicated ACK),  like TCP-DCR (Delayed Congestion Response TCP) \cite{leung07}.  

We introduce in the following subsections some standardized algorithms \cite{leung07} used by TCP to avoid packet reordering problems.

\begin{figure}[h]
\centering
\includegraphics[scale=0.5]{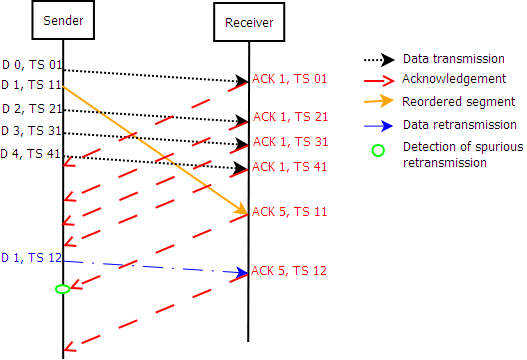} 
\caption{Eifel Algorithm}
\label{EiffelAlgo}
\end{figure}

\subsection{Eifel Algorithm}

Figure \ref{EiffelAlgo} illustrates how Eifel algorithm \cite{EifelRFC} \cite{Eifel} works. The sender inserts a TCP timestamp option in each transmitted segment, and the receiver inserts the timestamp value of the received segment in the corresponding acknowledgement. In case of loss, the sender saves the values of the current congestion window (cwnd) and of the slow start threshold (ssthresh). Then the sender retransmits the missing segment and stores its timestamp value. When the sender receives an acknowledgement for a retransmitted segment, it compares the saved timestamp value with the one inserted in the acknowledgement. If the first one is greater then the retransmission is considered spurious and the values of cwnd and ssthresh are restored.

\begin{figure}[h]
\centering
\includegraphics[scale=0.5]{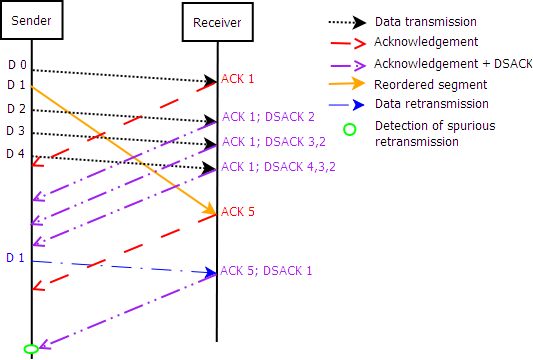} 
\caption{DSACK Algorithm}
\label{DsackAlgo}
\end{figure}

\subsection{DSACK Algorithm}

Figure \ref{DsackAlgo} illustrates how the DSACK algorithm \cite{DsackRFC} works. This algorithm is based on the SACK (Selective Acknowledgement) option. At the reception of a segment that creates a hole in the sequence numbers, the receiver sends back a duplicated acknowledgement containing a SACK option. The first block in the SACK option refers to the segment which triggers this duplicated acknowledgements. After three duplicated acknowledgements, the sender retransmits the missing segment, saves the congestion window value, and then enters a Congestion Avoidance phase. After that, when the sender detects that the retransmitted segment was acknowledged twice, it infers a spurious retransmission and begin a DSACK slow start to the stored congestion window's value.

\vspace{1cm}

\section{Implementation}\label{sec:implementation}

We have implemented MPTCP under the network simulator NS-3 \cite{ns3} to evaluate MPTCP's performance and to compare the different congestion control mechanisms proposed in \cite{raiciu09}, and also to analyze the efficiency of our own mechanisms, such as those presented in the next section. The project is hosted at Google Code \cite{MpTCP}. We opted for this simulator for its many interesting features. NS-3 is an open source network simulator, it is available for free under the GNU GPLv2 \cite{gpl} license. NS-3 has a large community of developers and users, and it is the main network simulator used in the academic area.

In the following subsections, we illustrate the different states a MPTCP connection can take, the structure of our MPTCP implementation, the segment flow through the different used classes and the implemented algorithms for packet reordering.

\subsection{Connection states}

\begin{figure}[h]
\centering
\includegraphics[scale=0.5]{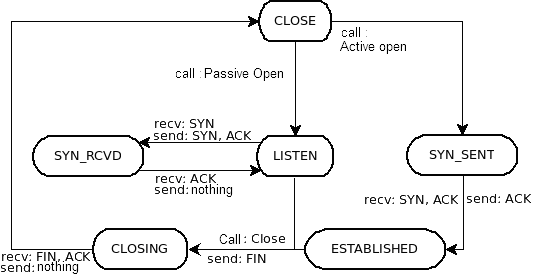} 
\caption{MPTCP connection's states}
\label{ConState}
\end{figure}

The MPTCP draft does not define for the moment any diagram to describe the different possible states of an MPTCP connection and their transitions. However, subflows may have the same states as a Standard TCP connection. For simplification, we have implemented a subset of the TCP states as described in figure \ref{ConState}. This figure illustrates the state diagram for a MPTCP subflow to exchange data, establish and close a connection. The starting and ending states are the state CLOSED.

The ESTABLISHED state allows data transfer between endpoints. Transitions to this state correspond to a connection opening, while transitions from it correspond to a connection closing on the corresponding subflow. 

To send data, the application on the source host asks the transition from CLOSE state to SYN-SENT state -- where the host has sent a SYN request ans is waiting for a SYN-ACK answer. When the source host receives a SYN-ACK segment, the connection moves to ESTABLISHED state and an ACK segment is sent. 

The connection on the destination host moves from CLOSE state to LISTEN state, after a passive opening by the application. When the destination receives a SYN segment, the connection moves from LISTEN to SYN-RCVD state and a SYN ACK segment is sent back to the source. 

The application can request to close the connection. The connection is closed by moving from any state to CLOSING state and sending a FIN segment. After the receipt of a FIN ACK segment, the connection is totally closed and moved to CLOSED state .

\subsection{Structure}

The multipath transport layer is divided into two sublayers. The upper sublayer is responsible for the connection management: establishing connection, initiating subflows, etc. It is called the MPTCP sub-layer. The lower sub-layer is responsible for the management of each sub-flow.

\begin{figure}[h]
\centering
\includegraphics[scale=0.75]{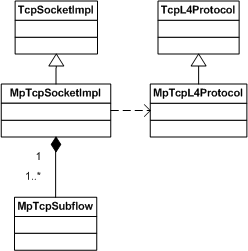} 
\caption{A subset of MPTCP classes}
\label{ClassDia}
\end{figure}

The following classes (figure \ref{ClassDia}) are the main ones composing the MPTCP transport layer: 
\begin{itemize}
\item \texttt{MpTcpSocketImpl} is a subclass of the NS-3 class \texttt{TcpSocketImpl}. It provides to the application layer a MPTCP API (connect, bind, etc.) to manage Multipath TCP connections. It also implements the packet reordering algorithms described previously.
\item \texttt{MpTcpL4Protocol} is a subclass of the NS-3 class \texttt{TcpL4Protocol}. It is an interface between the multipath transport layer and the network layer.
\item \texttt{MpTcpSubflow} represents a subflow of a MPTCP connection. 
\item \texttt{MpTcpHeader}  is a subclass of \texttt{TcpHeader}. An instance of this class is a TCP header that can handle options like TimeStamp, MPC, etc.
\end{itemize}

\subsection{Segment flow}

\begin{figure}[h]
\centering
\includegraphics[scale=0.5]{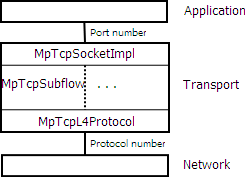} 
\caption{Multipath transport layer}
\label{MpTcpLayer}
\end{figure}

Figure \ref{MpTcpLayer} shows the structure of the multipath transport layer and its interfaces with the application and the network layers. The application and the multipath transport layers use the port number as an interface to communicate with each other. Similarly, the network and the multipath transport layers use the protocol number.

When the application has data to send, it chooses the appropriate transport layer instance using the peer source and destination port number. The \texttt{MpTcpSocketImpl} receives the data, splits it into TCP segments of a maximum size MSS \footnote{MSS: Maximum Segment Size} and then forwards it to the appropriate subflow which is represented by a \texttt{MpTcpSubflow} instance.
At this level, a TCP segment header is created and will contain one of the MPTCP options (DSN option if the segment contains data). After that, the segment is ready to be sent, it is forwarded to the \texttt{MpTcpL4Protocol} which will forward it to the network layer.

At the receiver side, when the network layer receives a packet, it figures out the corresponding protocol number and sends the segment to the appropriate transport instance. When the \texttt{MpTcpL4Protocol} instance receives the segment, it uses the network layer addresses to find the corresponding subflow and then forward the segment to a \texttt{MpTcpSubflow} instance. The \texttt{MpTcpSubflow} updates the information related to its subflow (ex: sequence number), then it sends the segment to the \texttt{MpTcpSocketImpl}. The later adds the data to the previously received ones, notifies the application about the reception of new data and optionally generates a response (ex: sends back an acknowledgement to the source).

\subsection{Discussion}
In NS-3, the transport layer is implemented via the \texttt{TcpSocketImpl} class. This class holds many variables like {\tt m\_endPoint} which is an instance of \texttt{Ipv4EndPoint}. This variable maintains information about a data flow (source - destination ports and addresses), and a callback for notification to higher layers that a packet from a lower layer was received.
 
When a packet is received by an instance of {\tt TcpL4Protocol}, this one uses a callback (method {\tt ForwardUp} of {\tt m\_endPoint}) to notify the upper layer (which is represented by {\tt TcpSocketImpl}) of the reception. In fact, {\tt TcpL4Protocol} holds a list of pointers to {\tt Ipv4EndPoint} variables each one attached to a {\tt TcpSocketImpl} instance.

In our MPTCP implementation, the classes composing the transport layer are derived from the ones of NS-3 ({\tt TcpSocketImpl, TcpL4Protocol}). 

The {\tt MpTcpSocketImpl} class maintains a set of subflows, a received packet may below to one of them and the {\tt m\_endPoint} variable is used to determine which one. 

When {\tt MpTcpL4Protocol} receives a packet, it updates the {\tt m\_endPoint} and notify the {\tt MpTcpSocketImpl} about the reception of the packet. 

The {\tt MpTcpL4Protocol} may receive at the same time a lot of packets belonging to different subflows. In this case, we expected that the packets will be forwarded up one by one which make easy the use of the {\tt m\_endPoint} variable -- it will contain the information of the right subflow. Instead of this, packets are forwarded up (after removing the IPv4 header) at once and not in the reception order. Thus, the {\tt m\_endPoint} variable will not contain a consistent information. 

Such a behaviour is not a problem for Standard TCP. For Multipath TCP it is because we can not decide to which subflow a packet belongs.

To overcome this, we used the sequence number in the segment header and for each subflow we gave a different sequence space.

\section{Simulations}\label{sec:simulation}
We used the implementation previously described to run a set of simulations in order to evaluate MPTCP connection performance by varying network parameters: bandwidth, latency and loss rate. Default values for these parameters are respectively: 0.5 Mb/s, 10 ms, 0. 
We also varied the used congestion control algorithm (Uncoupled TCPs, Linked Increases, etc.) and the packet reordering algorithm (none, Eifel, DSACK). 

\begin{figure}[h]
\centering
\includegraphics[scale=0.5]{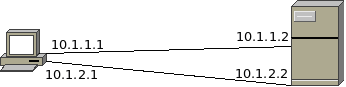} 
\caption{Simulated system}
\label{simulation}
\end{figure} 

The simulated system (fig. \ref{simulation}) is composed of a FTP application, to transfer a 10 Go file, running on Client/Server architecture where the two hosts are linked by two point to point links.

\begin{figure}
\centering
\includegraphics[scale=0.25]{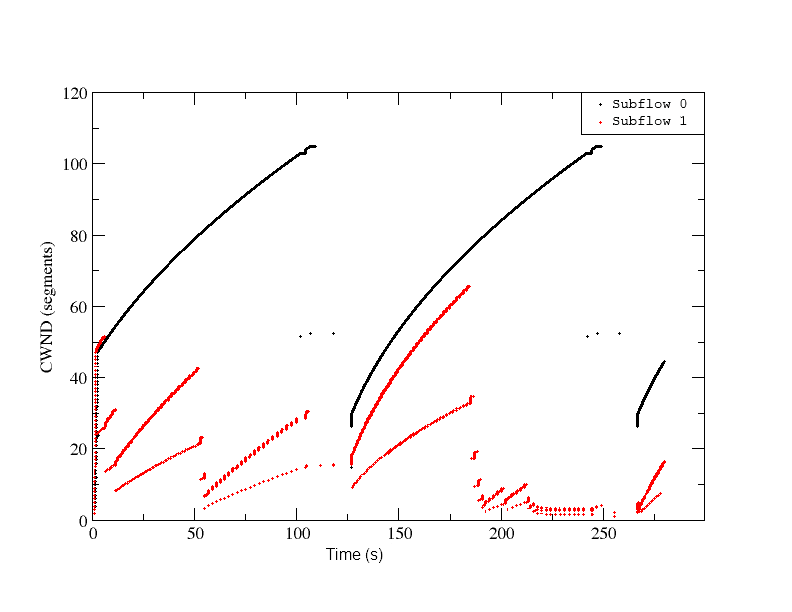} 
\caption{Congestion window evolution in case Eifel is used}
\label{cwndEifel}
\end{figure}

Figure \ref{cwndEifel} shows the behaviour of the congestion window in case the Eifel algorithm is used. The graph of the congestion window of the subflow 1 oscillates between two curves of evolution: the congestion window takes its values according to the lower one in case of segment retransmission, and returns to the upper one when the Eifel algorithm detects that the retransmission was spurious.

\begin{figure}
\centering
\includegraphics[scale=0.25]{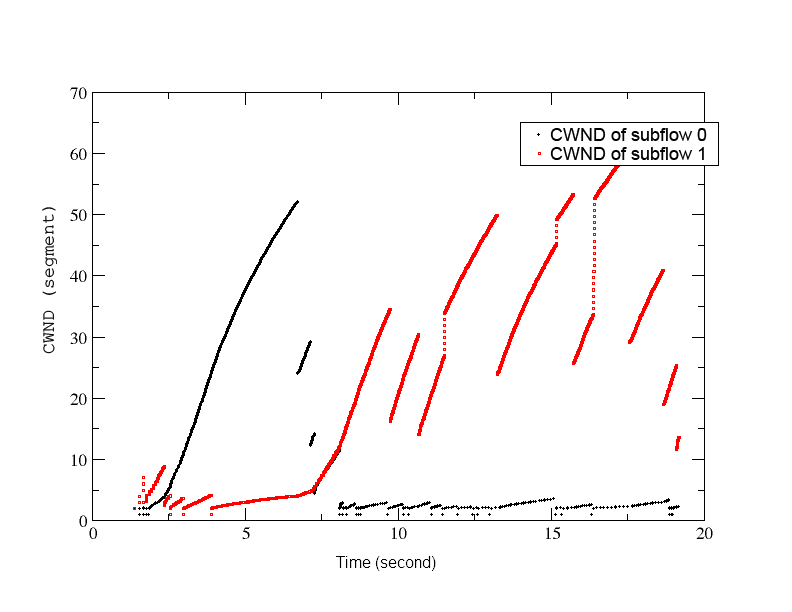} 
\caption{Congestion window evolution in case DSACK is used}
\label{cwndDsack}
\end{figure}

Figure \ref{cwndDsack} illustrates the evolution of the congestion window of two subflows of a MPTCP connection in case the DSACK algorithm is used. In the graph of the congestion window of subflow 1, we can see three periods of time during which the corresponding window grows unusually in an exponential way. These periods reflect the DSACK Slow Start which is triggered by the DSACK algorithm after detecting a spurious retransmission.

\section{Conclusion}\label{sec:conclusion}

We have implemented the MPTCP protocol under the network simulator NS-3. The implementation is conform to the IETF drafts, at least according to their release in their July 2010. We also have added for this implementation of MPTCP two packet reordering mechanisms standardized for TCP.

We use now this implementation to test on simulations nonetheless under different realistic environments various congestion control and packet reordering mechanisms that could be used with the MPTCP protocol. Our objective to evaluate and compare the robustness and the performance of these different congestion control and packet reordering mechanisms.  

Based on the conducted simulations, we deduced that neither Eifel nor DSACK will help to face effectively performance problems due to persistent out-of-sequence packets arrival. We plan to test other packet reordering mechanisms and assess their impact on avoiding or at least alleviating this problem.

\end{document}